 \documentclass[preprint2]{aastex}
 \usepackage{epsfig}

\newcommand{\etal}{\textit{et al.}\ }

 \slugcomment{ \textit{Icarus}, in press. }

\shorttitle{Titan's Atmosphere in Late Southern Spring}
\shortauthors{Roe \et}

\begin{document}


\title{
NOTE: Titan's Atmosphere 
in Late Southern Spring  \\ Observed with Adaptive
Optics on the W.\ M.\ Keck II 10-meter Telescope\altaffilmark{1}}


\author{\vskip -0.3in
Henry G.\ Roe,\altaffilmark{2} 
Imke de Pater,\altaffilmark{2} 
Bruce A.\ Macintosh,\altaffilmark{3}
Seran G.\ Gibbard,\altaffilmark{3}
Claire E.\ Max,\altaffilmark{3}
\and
Chris P.\ McKay\altaffilmark{4}
\\
\vspace{0.1in}
\textbf{\textit{Icarus}, in press.}
\vspace{-0.4in}
}


\altaffiltext{1}{Data presented herein were obtained
at the W.M.\ Keck Observatory, which is operated as a scientific 
partnership among the California Institute of Technology, the 
University of California, and the National Aeronautics and Space
Administration.  The Observatory was made possible by the generous
financial support of the W.M.\ Keck Foundation.}
\altaffiltext{2}{Department of Astronomy, 601 Campbell Hall, University
of California, Berkeley, CA 94720-3411.  Email address of the 
corresponding author: hroe@astro.berkeley.edu}
\altaffiltext{3}{Lawrence Livermore National Laboratory, Livermore, CA 94550}
\altaffiltext{4}{NASA Ames Research Center, Moffett Field, CA 94035}


\begin{abstract} 
\baselineskip=10pt
Using adaptive optics on the W.M.\ Keck II telescope we 
imaged Titan several times during 1999 to 2001 in
narrowband near-infrared filters selected to probe Titan's
stratosphere and upper troposphere.  We observed a bright feature around
the south pole, possibly a collar of haze or clouds.  Further, we find
that solar phase angle explains most of the observed east-west brightness
asymmetry of Titan's atmosphere, although the data do not preclude the 
presence of a `morning fog' effect at small solar phase angle.
\end{abstract}











\baselineskip=12pt
\textheight 9.5in
\topmargin -1.0in
\section{Introduction}

\begin{deluxetable}{ccccccc}
\tabletypesize{\scriptsize}
\tablecaption{Parameters of Titan Observations. \label{tbl-1}}
\tablewidth{0pt}
\tablehead{
\colhead{} & \colhead{} & \colhead{} & \colhead{} & \colhead{} & \colhead{Sub-Earth} & \colhead{Solar} \\
\colhead{UT} & \colhead{UT} & \colhead{Filter} & \colhead{Exposure} & \colhead{Apparent} & \colhead{Latitude,} & \colhead{Phase Angle,} \\
\colhead{Date} & \colhead{Time} & \colhead{} & \colhead{Time} & \colhead{Size\tablenotemark{a}} & \colhead{Longitude\tablenotemark{a,b}} & \colhead{Position Angle\tablenotemark{a,c}} \\
}
\startdata
30 October 1999 & 11:00 & J1158 & 3$\times$120 sec & 0$\farcs$865 &
-19$\fdg$92, 108$\fdg$88 & 0$\fdg$8870, 91$\fdg$42 \\
30 October 1999 & 11:36 & H1702 & 3$\times$120 sec & 0$\farcs$865 &
-19$\fdg$92, 109$\fdg$45 & 0$\fdg$8842, 91$\fdg$48 \\
17 August 2000 & 15:36 & J1158 & 4$\times$120 sec & 0$\farcs$774 &
-24$\fdg$05, 208$\fdg$84 & 6$\fdg$3375, 77$\fdg$86 \\
17 August 2000 & 15:13 & H1702 & 4$\times$120 sec & 0$\farcs$774 &
-24$\fdg$05, 208$\fdg$48 & 6$\fdg$3374, 77$\fdg$85 \\
20 February 2001 & 6:24 & J1158 & 4$\times$120 sec & 0$\farcs$774 &
-23$\fdg$14, 108$\fdg$14 & 6$\fdg$1893, 256$\fdg$35 \\
20 February 2001 & 5:50 & H1702 & 4$\times$120 sec & 0$\farcs$774 &
-23$\fdg$14, 107$\fdg$60 & 6$\fdg$1895, 256$\fdg$35 \\

 \enddata


\tablenotetext{a}{Obtained from the JPL Horizons Ephemeris, available
at http://ssd.jpl.nasa.gov/horizons.html}
\tablenotetext{b}{Latitude and longitude coordinates are planetographic.}
\tablenotetext{c}{These coordinates describe the position of the sub-Solar 
point relative to the sub-Earth point on Titan.  The phase angle
is defined as the Sun-Titan-Earth angle.  The position angle (PA)
is defined CCW with respect to direction of the true-of-date 
Celestial North Pole.}


\end{deluxetable}

Previous high-angular resolution studies of Titan have focused
for the most part on mapping surface albedo  
\citep{meier2000,gibbard1999,combes1997,smith1996,coustenis2001}.
Several of these works derived atmospheric parameters,
such as haze opacity, in order to separate the atmospheric and
surface contributions.  Most recently, \citet{coustenis2001} used 
moderate bandwidth filters ($\sim0.15 \mu$m FWHM)
 in the J- and H-bands on the 3.6 m Canada-France-Hawaii Telescope to
image the surface and detected increased limb brightening around Titan's
South Pole and on Titan's morning limb.  Titan's morning limb is
to the East on the sky as viewed from Earth.  \citet{coustenis2001}'s 
observations were just past Titan opposition from Earth with a 
Sun-Titan-Earth angle of only 0$\fdg$5.  If, as one might expect,
solar phase angle determines the East-West asymmetry in limb-brightening,
then \citet{coustenis2001} would have seen increased limb brightening
on Titan's evening limb.  They interpreted this discrepancy from
expectation by suggesting that a `morning fog' exists on Titan due
to small diurnal changes in the thermal structure of the atmosphere and
condensation of ethane as a parcel of air moves from the dark side
of Titan to the sunlit side.  The data we present here show that
at larger solar phase angles, solar phase angle dominates the
East-West limb-brightening asymmetry; however, our results are consistent
with a `morning fog' which would determine the
East-West limb-brightening asymmetry at smaller phase angles.
We also resolve the brightening seen at the southern limb by 
\citet{coustenis2001} and \citet{meier2000}
into a `collar' around the south pole at a 
planetographic latitude of 70$\fdg$S to 75$\fdg$S.

Since October 1999 we have imaged Titan 
using the adaptive optics (AO) system on the W.M.\ Keck II telescope
\citep{wizinowich2000}
in several narrowband filters.  In this Note we present the data from 
narrowband filters centered at 1.158 $\mu$m and 1.702 $\mu$m, 
chosen to probe only the upper atmosphere of Titan above the 
middle-to-upper troposphere ($>\sim 20$ km altitude).  The contribution
from surface reflection is negligible; essentially all the light we
observe in these filters is sunlight scattered by particles in the
atmospheric haze layers \citep{lemmon1995}.
These images show a number of features in Titan's atmosphere: 
a bright collar around Titan's south pole, 
northern polar limb-brightening, and an east-west assymmetry in
limb-brightening.  

\section{Observations}

The images presented in this Note were obtained on the dates listed
in Table 1.  The October 1999 data were taken 7.1 days before opposition,
with a Sun-Titan-Earth angle of $\sim0\fdg9$.  The August 2000 and
February 2001 data are from opposite sides of opposition and the solar
phase angle is nearly equal in size ($\sim6\fdg$) 
for these two data sets.
Further  details of the observations are listed in Table 1.

The data were collected using two different instruments.  The October 1999
data were collected with
KCAM, the Keck Observatory's initial camera for use with Adaptive
Optics (AO).   
KCAM contains a 256$\times$256 NICMOS-3 array with a platescale of
0.0175''/pixel.  Later data were collected using the slit-viewing
camera SCAM of the observatory's near-infrared spectrometer NIRSPEC
\citep{mclean1998}.
SCAM is a 256$\times$256 Rockwell HgCdTe array.  When NIRSPEC is used behind
the AO system a set of warm reimaging optics are used to
form a platescale of 0.017''/pixel on SCAM.  In spite of the warm 
reimaging optics in front of SCAM, data from SCAM are of higher quality
than data from KCAM due to a higher quality array, better optics, 
lower noise in the electronics, and a better baffled optical design with
a true cold pupil stop.  Between the AO bench and either instrument
sits a room-temperature filter wheel, in which the narrowband filters used
in the current work reside.

The filters used in the current work are IR bandpass filters purchased from 
Coherent\footnote{http://www.coherentinc.com/}.  The H1702 filter has
 a maximum transmission of 0.48 at 
a central wavelength of 1.702 $\mu$m with a full-width-half-max (FWHM)
of 0.017 $\mu$m, and the J1158 filter has a maximum transmission of 
0.58 at 1.1582 $\mu$m with a FWHM of 0.0175 $\mu$m.  
The appropriate internal cryogenic J- or H-band filter of NIRSPEC or
KCAM is used to block any short or long wavelength light leaks of
the J1158 and H1702 filters.  

\begin{figure}
\vskip 0in
\hskip -0.5in
\epsfig{figure=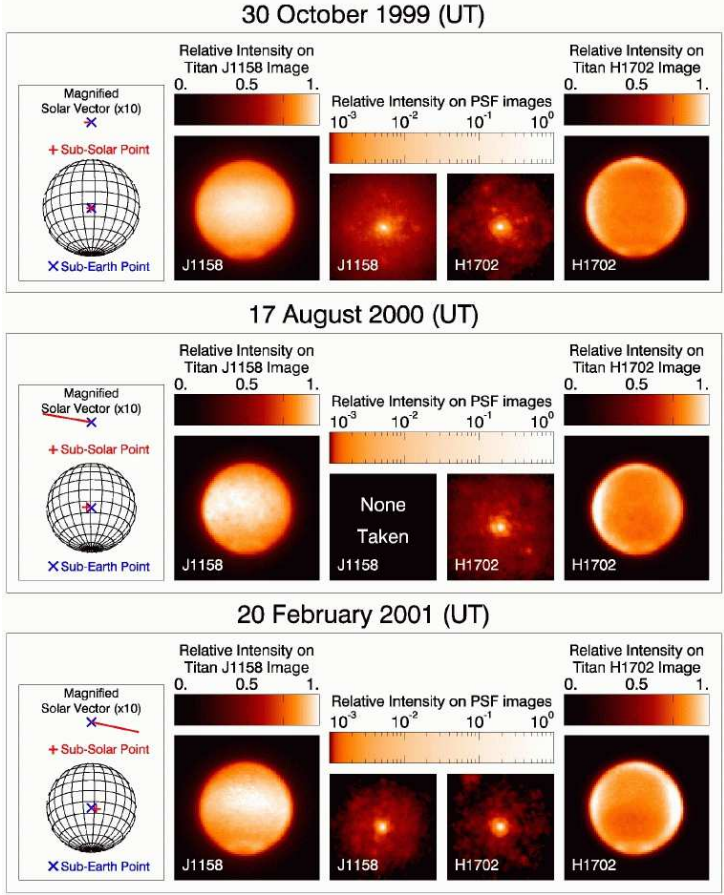,width=3.8in,angle=0}
\vskip 0in
\caption{\small
Images of Titan and PSF star.  Each horizontal panel
shows the observations from a different date.  At left-most in
each panel is a line figure showing the orientation of Titan's
disk and location of the sub-solar and sub-Earth points on Titan in
the images for that date.  Titan's solid radius is 2575 km.
The line figure of Titan's orientation 
is shown at the same scale as Titan, assuming an apparent radius 
of 2650 km to partially account for the extent of Titan's atmosphere, 
with lines of latitude and longitude plotted every $15\fdg$. 
 Above this line figure is a graphic showing the direction
and size of the vector from the sub-Earth point to the sub-solar point
magnified by a factor of 10.  The Titan J1158 images are all shown on
the same linear color scale, as indicated by the color bars above the
images.  The Titan J1158 images were normalized such that the average
intensity within the central $\pm4$ pixels is the same in all three
images.  Similarly, the Titan H1702 images are all shown on the same
color scale, and were normalized such that the average intensity within
the central $\pm6$ pixels is the same in all three images.
The PSF images are all shown on logarithmic color scales
relative to the peak intensity in each image.  
\label{fig1}}
\end{figure}

Total exposure times for each image of Titan shown in Fig.\ 1
are listed in Table 1.  Each night's observations included a minimum
of 3 images of Titan in different positions on the array to minimize
the effect of bad pixels and noise in the flat-field correction.  The
sky background in both filters was observed to be extremely low ($\ll1 $
ct pixel$^{-1}$ sec$^{-1}$),
in spite of each filter's overlap of several night-sky OH lines 
\citep{rousselot2000}.  In some cases a separate sky frame at the same
exposure length was obtained by offseting 10-20'' from Titan.  In cases
without a separate sky frame, dark frames from the beginning or end of
the night were used to construct a median dark, which was subtracted
from the images of Titan.  Correction for the variations in the
flat-field or pixel-to-pixel sensitivity were made using data taken
on the twilight sky in the broadband H filter without the H1702 filter,
and in the broadband J filter without the J1158 filter.
On several evenings we also took these flat-field frames with the H1702
filter.  These flat-field data taken with the H1702 filter show 
a decrease in the signal-to-noise ratio from the wideband H flat-field data
but no other significant differences.  Similarly we found no significant
difference between flat fields observed in the J-band with or without the
J1158 filter.

After sky or dark subtraction and flat-field correction, the 
permanently bad pixels in each image are replaced with the median
of the nearest 4 to 8 good pixels.  Each 
image of Titan is then cross-correlated to every other image
of Titan from the same evening in the same filter.  
Our cross-correlation algorithm sparsely
samples the (x-y)-lag plane and homes in on the 
peak of the cross-correlation
function with a near-minimum of computing time.  A second step of the
algorithm is to then rebin each image by a factor of 4 (16 new pixels
per 1 original pixel) and find the peak of the 
cross-correlation function to 1/4 pixel precision.  
We then do a linear-least squares fit to this matrix
of offsets to find the shifts between all the images.  
The bad pixels in each image are remasked with zeroes before the final 
step of using bilinear interpolation to shift the images.  Each pixel
in the final coadded image is divided by the total exposure time it 
represents, which varies due to the masking of bad pixels.
The final step of image processing was to rotate
each image by the instrument position angle and Titan's N-S position
angle on the sky, such that Titan's N-S axis aligns with the y-axis of the 
images shown in Figure 1.  In the case of KCAM images, a flip about the 
vertical axis is necessary before this rotation in order to correct
to the astronomical standard of sky north-up and sky east-left.  
Thus, the images
of Titan in Fig.\ 1 are presented as Titan would normally be seen 
on the sky with Titan's north-south axis aligned vertically.

In almost all cases a star of approximately the same visual magnitude
as Titan was also observed before
or after Titan in order to monitor the point spread function (PSF) of
the AO correction.
Processing of the PSF data is identical to the processing of 
the Titan data as described above, including the final step of image
rotation so that the PSF images in Fig.\ 1 are at the same orientation
as the corresponding images of Titan.  Images of the PSF stars are
shown on a logarithmic color scale in order to better display the wings
of the PSF.  The total exposure time each PSF image contains ranges from
10 to 100 sec.  All data presented here were taken on nights of average
to better-than-average seeing ($\le\sim 0\farcs5$).  Images of the PSF stars
show that the FWHM of the AO-corrected PSF was 45 to 48 milliarcsec in
J1158 and 47 to 53 milliarcsec in H1702, both well sampled by our
pixels of roughly 17 milliarcsec.  These are within a factor of
two of the theoretical limit for a 10-meter telescope (1.02 $\lambda/D = $
24 milliarcsec at 1.158 $\mu$m and 36 milliarcsec at 1.702 $\mu$m).
A PSF with a FWHM of 50 milliarcsec means that the spatial resolution
of our data is approximately 300 km referenced to Titan's surface.

\section{Atmospheric Calculations}

\begin{figure}[t]
\vskip  -0.0in
\hskip 0in
\epsfig{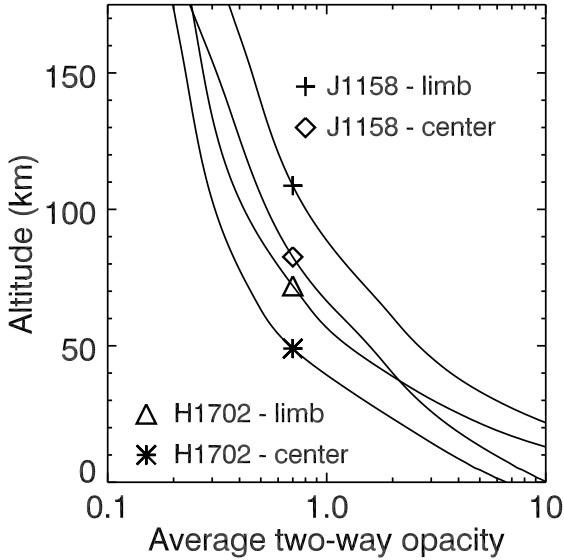}
\vskip 0in
\caption{\small
Altitudes probed by each filter.  The y-axis is the 
altitude above Titan's surface and the x-axis gives the average roundtrip
methane opacity as a function of altitude as calculated using 
Eq.~\ref{eqn:avgtau}.  
Plotted are curves for both J1158 and H1702 filters at two different viewing 
points on Titan.  The `center' geometry is for a line of sight pointed
straight at the center of Titan's disk.  The `limb' geometry is for
 a line-of-sight at Titan's limb that will cross just tangent to the solid
surface.  See text for more details on the definition of `average
opacity' in a filter.  This figure shows that the H1702 filter probes
several tens of kilometers deeper than the J1158 filter.  Neither
filter significantly probes the surface.  
\label{fig2}}
\end{figure}

While an attempt to fit these images with a scattering/absorbing 
radiative transfer model of Titan's atmosphere is beyond the scope of
this short Note, estimating the altitudes probed by our narrowband filters
is useful for initial interpretation.  
Figure 2 shows the average roundtrip methane opacity encountered by
a photon as a function of altitude.  For each filter two cases are included.
The first case is for a line-of-sight directed at the center of Titan's 
disk.  The second case is for a line-of-sight directed at Titan's
solid-surface limb.  Our simple calculations ignore opacity due to
haze scattering since at the wavelengths of our filters  opacity due to
methane is much greater than the opacity from all other sources.
We use the profile of \citet{yelle} for the
atmospheric structure and methane abundance, and assume
that each layer is a spherical shell.  The methane
abundance is 3$\%$ at the surface and in the troposphere, elsewhere
the methane abundance is set to saturation.  We use the
self-broadened methane \textit{k}-coefficients of \citet{irwin}
for these opacity calculations.  

A more complete description of the 
correlated-\textit{k} method is given in \citet{thomasandstamnes}, but
it is worth making explicit here what we mean by `average opacity' for a
filter.  The \citet{irwin} \textit{k}-coefficients are calculated for 
5 cm$^{-1}$ wide bins, and the bandwidths of both of our filters 
are covered by several tens of these wavenumber bins.  Our model
atmosphere contains 326 layers which we number from 
$l = 0$ at Titan's surface to $l = L_{top}$ at an altitude of 1300 km.  
Layers at altitudes $>\sim 500$ km are essentially
irrelevant for these current calculations.
In wavenumber bin $w$ the average 
roundtrip transmission from outside the atmosphere to the 
bottom of layer $L$ and back out of the atmosphere is
\begin{equation}
  T(w,L) = \sum_{i=0}^{9} g_{i} \exp\left[ - \sum_{l=L}^{L_{top}} 2 k_{i,l}
           u_{l} \right],
\end{equation}
where: $g_{i}$ is the weight of the $i$'th Gaussian quadrature point of the
\textit{k}-distribution calculation, the abundance of methane 
$u_{l}$ is calculated assuming that layer $l$ is a spherical shell and
that the atmosphere in layer $l$ is homogeneous, and 
$k_{i,l}$ is the \citet{irwin} $k$-coefficient for the $i$'th 
Gaussian quadrature point interpolated/extrapolated for the temperature
and pressure of the $l$'th atmospheric layer.  The factor of 2 is 
included because we are interested in the roundtrip path.
The `average' opacity in a filter $\tau(L)$ is then
\begin{equation}
 \tau(L) = - \ln\left( \sum_{w=1}^{W} f_{w} T(w,L) \right)
\label{eqn:avgtau}
\end{equation}
where $f_{w}$ is the relative transmissivity of the filter in each
wavelength bin, normalized such that $\sum_{w=1}^{W} f_{w} = 1.0$.
For the two filters J1158 and H1702 the average opacity of Eq.~\ref{eqn:avgtau}
is shown in Fig.~\ref{fig2}.  

\section{Observed Features: East-West Asymmetry}

As discussed earlier, previous observers noted an 
asymmetry in the limb-brightening between the East/West or 
morning/evening limbs they could not attribute to phase angle effects.  
\citet{coustenis2001} interpreted this as 
evidence for a `morning fog' effect.  Our images of Titan are displayed
in the usual convention of Titan north up, with east-west displayed
as they appear on the sky.  Thus, given Titan's prograde rotation,
the morning limb is on the left in our images.  If no winds exist the
atmosphere at the morning limb is emerging from $\sim$190 hours of
darkness.  Given recent measurements of Titan's zonal winds \citep{kostiuk},
this period of darkness experienced by an atmospheric element could
be as short as $\sim$70 hours.  As described here, an initial
interpretation of our data is consistent with, but does not require,
a brightening of the morning limb beyond what would be expected due to 
phase angle considerations.

Our August 2000 and February 2001 images show obvious east-west
asymmetry in both the J1158 and H1702 filters.  The asymmetry 
appears almost exactly reversed between these two dates of
observation.  As shown graphically in the left-most panels of Fig.\ 1,
the solar phase angle is of almost exactly equal size ($\sim6\fdg$)
and opposite direction on these two dates.  In both cases the 
asymmetry is brighter in the direction of the sub-solar point, as
would be expected if the asymmetry is to be explained by the solar phase
angle.

The October 1999 data were taken at a much smaller solar phase angle
($0\fdg9$).  In this case the J1158 image shows essentially no east-west
asymmetry.  On the other hand, the H1702 image exhibits an east-west
asymmetry and the sign of the asymmetry is in agreement with the 
direction of the solar angle.  Since these observations were taken
before opposition this brightening is on the morning limb.  Therefore,
without more detailed analysis and modeling, these observations neither
confirm nor rule out the presence of a `morning fog.'

We therefore conclude that our data show definitively that at
larger solar phase angles the effect of the solar phase angle
dominates the observed east-west asymmetry, but that our data
do not rule out the `morning fog' hypothesis of  \citet{coustenis2001} 
at small phase angles, but also do not appear to require such a 
hypothesis.

\section{Observed Features: Southern Polar Collar 
and Northern Limb Brightening}

All of the images presented here show a bright feature about the south pole
and the H1702 images show 
a brightening at the Northern limb.  This southern feature appears to be a
bright collar with a hole centered on the south pole, rather than
a cap covering the entire polar region.  In the North-South direction
the collar appears to be unresolved ($<\sim 270$ km 
projected on our images, or 
$<\sim 380$ km referenced to Titan's surface at this latitude) and centered
between $70\fdg$S and $75\fdg$S latitude.  The J1158 images are typically
limb-darkened, while the H1702 images are limb-brightened.  A simplistic
explanation for this difference can be understood by realizing that
the light we are seeing is scattered off haze and that the bulk of this haze
sits in the lower stratosphere where the opacity due to methane is 
significantly different at J1158 wavelengths versus H1702 wavelengths.  At
J1158 wavelengths the haze is embedded in greater methane opacity 
compared to H1702 wavelengths, leading to the observed difference in
limb brightening and darkening.

From the transmission calculations presented in Fig.~\ref{fig2} it
is clear that these features are atmospheric, rather than lying on
Titan's surface.  Using the the average transmission curves of
Fig.~\ref{fig2} we can draw some conclusions about the altitudes of
these northern and southern features.  The northern limb brightening
is apparent in H1702 images, but is missing from J1158 images.  The 
absence of this feature from the J1158 data indicates that this
northern haze or cloud must lie below approximately 90 km altitude,
where the `J1158-limb' opacity is equal to unity.  Similarly, the 
presence of the northern brightening in the H1702 data leads us to
conclude that this northern haze or cloud sits more than 20-30 km
above Titan's surface.  Titan's tropopause is at approximately 40 km
altitude.  Therefore, our data constrain this northern atmospheric
feature, at the time of the observations, to the lower stratosphere or
upper troposphere.

Similar arguments can be made about the southern polar collar, which 
is apparent in both J1158 and H1702 images.  No upper limit on the altitude
of the haze or cloud forming this polar collar can be easily deduced without
further radiative transfer modeling and data analysis.  However,
the presence of the collar in the J1158 data lead us to conclude that 
this collar of material must be at or above 40 to 50 km altitude.  
Thus, our data and calculations restrict the
southern polar collar to be above the tropopause and at higher altitudes
than the cloud or haze observed on the northern limb.

We suggest that these Northern and Southern features are of a seasonal
origin.  Voyager II in visible light saw a dark Northern polar collar
or hood in early Northern Spring \citep{smith1982}, 
while our observations are in late Southern Spring.  It is possible that
Voyager II's Northern polar hood is of the same seasonal
origin as the Southern Polar collar we observed.  A plausible explanation,
inspired by an investigation of C$_4$N$_2$ by \citet{samuelson}, is
that during the long dark winter of Titan's poles significant gas-phase
concentrations of reaction products such as C$_4$N$_2$ accumulate.
At or above super-saturated concentrations particles will condense out.
Temperature in the polar stratosphere appears to lag behind
 insolation, such that during spring the polar stratosphere is
actually cooling \citep{bezard1995}.
Thus, during spring as the polar stratosphere cools
the rate of condensation of these particles is enhanced.  Assuming some
combination of: continued condensation of new particles into late spring
as the polar atmosphere reaches its annual minimum temperature, with
particle fall-out times of $>\sim$1 Earth year, we would expect to see
the remains of this polar winter buildup through late spring.  Such
a phenomenon matches what we observe near Titan's south pole.  Further,
a large buildup of haze particles over the northern 
winter pole could be expected to extend to latitudes just south of
the fully shadowed polar region, giving an explanation for the bright
feature we see at Titan's northern limb. 

In this hypothesis for explaining the southern polar collar 
we would expect the collar to fade in prominence and possibly
disappear as spring progresses towards the southern summer equinox in late
2002.  From October 1999 to February 2001 the southern
feature does not appear to change significantly.  To better understand the
seasonal cycles on Titan we must continue to observe.  We intend to
continue monitoring Titan's atmosphere using this relatively new
technique of adaptive optics on large telescopes through the arrival
of the Cassini mission in 2004 and beyond.

\noindent
\textbf{Acknowledgements}

 We thank the staff of
the W.M. Keck Observatory, especially the members of the adaptive
optics team for all their hard work.
H.G.R.\ acknowledges support from a NASA GSRP grant funded through
NASA Ames Research Center and a
Sigma Xi Grant-in-Aid-of-Research from the National Academy of Sciences,
through Sigma Xi, The Scientific Research Society.
This work was performed under the auspices of the U.S. 
    Department of Energy, National Nuclear Security Administration 
    by the University of California, Lawrence Livermore National 
    Laboratory under contract No. W-7405-Eng-48.
This work has been supported in part by the National Science 
Foundation Science and Technology Center for Adaptive Optics, managed 
by the University of California at Santa Cruz under cooperative 
agreement No. AST-9876783.
The authors wish to extend special thanks to those of Hawaiian ancestry 
on whose sacred mountain we are privileged to be guests.  Without their 
generous hospitality, none of the observations
presented herein would have been possible.

\end{document}